\documentstyle[12pt,titlepage,epsfig]{article}
\font\medio=cmr10 scaled \magstep2
\outer\def\beginsection#1\par{\medbreak\bigskip
      \message{#1}\leftline{\bf#1}\nobreak\medskip
\vskip-\parskip
      \noindent}

\def\laq{\raise 0.4ex\hbox{$<$}\kern -0.8em\lower 0.62
ex\hbox{$\sim$}}
\def\gaq{\raise 0.4ex\hbox{$>$}\kern -0.7em\lower 0.62
ex\hbox{$\sim$}}

\def\beq{\begin{equation}}
\def\eeq{\end{equation}}
\def\bea{\begin{eqnarray}}
\def\eea{\end{eqnarray}}

\def \pa {\partial}
\def \ra {\rightarrow}

\def \pr {\prime}
\def \se {\prime \prime}

\def \la {\lambda}

\def \La {\Lambda}

\def \b {\beta}
\def \a {\alpha}
\def \ap {\alpha^{\prime}}

\def \Ga {\Gamma}
\def \ga {\gamma}

\def \da {\delta}

\def \noi {\noindent}

\begin{document}
\bibliographystyle {unsrt}

\titlepage
\begin{flushright}
CERN-TH/97-41 \\
hep-th/9707053\\
\end{flushright}
\vspace{12mm}
\begin{center}
{\bf A CLASS OF NON-SINGULAR GRAVI-DILATON BACKGROUNDS
\footnote{{\sl This essay received an ``Honorable Mention'' from
the Gravity Research Foundation, 1997 --- Ed.}}}

\vspace{10mm}

A. Buonanno${}^{(a,b)}$, M. Gasperini${}^{(c)}$ and C.
Ungarelli${}^{(a,b)}$\\ 
\vspace{6mm}

${}^{(a)}$
{\sl Theory Division, CERN, CH-1211 Geneva 23, Switzerland} \\
${}^{(b)}$
{\sl Istituto Nazionale di Fisica Nucleare, Sezione di Pisa, Pisa, 
Italy} \\
${}^{(c)}$
{\sl Dipartimento di Fisica Teorica, Universit\`a di Torino, \\
Via P. Giuria 1, 10125 Turin, Italy }\\
\end{center}
\vskip 1cm
\centerline{\medio  Abstract}

\noindent
We present a class of static, spherically symmetric,
non-singular solutions of the tree-level string effective
action, truncated to first order in $\alpha'$.  In the 
string frame the solutions  approach asymptotically (at $r\ra
0$ and $r\ra \infty$) two different anti-de Sitter
configurations, thus interpolating between two maximally
symmetric states of  different constant curvature. The
radial-dependent dilaton defines a string coupling which is
everywhere finite, with a peak value  that can be chosen
arbitrarily small so as to neglect  quantum-loop corrections. 
This example stresses the possible importance of finite-size
$\ap$ corrections, typical of string theory, in avoiding
space-time singularities.

\vspace{13mm}
\begin{center}
{\sl To appear in {\bf Mod. Phys. Lett. A}}
\end{center}
\vspace{5mm}

\vfill
\begin{flushleft}
CERN-TH/97-41\\
March 1997 
\end{flushleft}

\newpage

It has often been conjectured, soon after the appearance of
the first paper \cite{1} discussing string corrections to the
Schwarzschild metric, that the higher-derivative terms
(the so-called $\ap$ corrections), appearing at next-to-leading
order in the string effective action, should regularize the
curvature singularity present at the origin in the
Schwarzschild solution \cite{2,3}. It has been shown, in
particular, that the singularity may indeed disappear when the
delta-function of the effective point-like source, supporting
the solution at the origin, is smeared out by the effect of 
$\ap$ corrections \cite{4}. 

To support the expected ``smoothing" of short-distance
divergences, due to the fundamental cut-off scale  of
string theory, we will discuss in this paper static and
spherically symmetric solutions of the string effective action
in vacuum, with no sources (nor cosmological term) included in
the action, but with higher-derivative terms included up to
first order in $\ap$

In general relativity it is well known that all non-trivial
spherically symmetric solutions in vacuum are singular at the
origin, and that the only solution that is regular everywhere is the
trivial Minkowski manifold. For the lowest order string
effective  action all the non-trivial solutions are also singular.
In this paper we will show that, when we add
higher-derivative corrections to first order in $\ap$, the action
admit instead non-trivial solutions  in which the
curvature is bounded everywhere, and asymptotically
approaches two constant  finite values at $r\ra 0$ and $r \ra
\infty$. 

We shall work in the string frame, where the gravi-dilaton
effective action of critical string theory, at tree level in the
string-loop expansion but including first-order $\ap$
corrections, can be written in the form \cite{5}: 
\beq
S=\int d^{4}x \sqrt{-g}e^{-\phi} \left[-R-
(\nabla \phi)^2+{k\ap \over 4} \left(R^2_{GB} - 
(\nabla \phi)^4\right)\right] .
\label{1}
\eeq
Here $\phi$ is the dilaton field,  $R^2_{GB} 
\equiv R_{\mu\nu\a\b}^2-4  R_{\mu\nu}^2+
R^2$ is the Gauss--Bonnet invariant, and $k=1, 1/2$ for the
bosonic and heterotic string, respectively (conventions: 
$g_{\mu \nu}=(+---)$, $R_{\mu\nu\a}\,^\b = 
\pa_\mu \Ga_{\nu\a}\,^\b - ... ,$ and $R_{\nu \a}=
 R_{\mu\nu\a}\,^\mu$).  Note that the particular
field redefinition that we have chosen eliminates
higher-than-second derivatives from the field equations, but
necessarily introduces dilaton-dependent $\ap$ corrections in
the effective action (the Gauss--Bonnet term, by itself, may
represent the complete first-order $\ap$ corrections only in
the conformally related Einstein frame \cite{5}).

Looking for static and spherically symmetric solutions,
parametrized by 
\bea
&&
ds^2= e^\nu dt^2 - e^\la dr^2- {\cal R}^2 \left(d\theta^2+
\sin^2 \theta d \varphi^2 \right) , \nonumber\\
&&
\nu= \nu (r) ,  \,\,\,\,\,\, \la= \la (r) , \,\,\,\,\,\,
{\cal R}= {\cal R} (r) , \,\,\,\,\,\, \phi = \phi (r) , 
\label {2}
\eea
the effective action (after integration by parts) can be
rewritten as:
\bea
&&
S=4\pi \int dr e^{{\nu\over 2}-{\la\over 2}-\phi} \left[
{\cal R}^2 \phi^{\pr 2} -{\cal R}^2\nu'\phi' -4{\cal R} {\cal R}'
\phi' +2{\cal R}^{\pr 2}+ 2{\cal R} {\cal R}'\nu' +2 e^\la + \right.
\nonumber \\
&&
\left. + k\ap \nu' \phi' \left(e^{-\la}{\cal R}^{\pr 2}- 1\right) -
k {\ap\over 4}e^{-\la}{\cal R}^2\phi^{\pr 4} \right]
\label {3}
\eea
(a prime denotes differentiation with respect to $r$). By
varying $\la, \nu, {\cal R}$ and $\phi$, and imposing the radial
gauge ${\cal R}=r$, we obtain respectively the equations (we
put $k=1$ for simplicity)
\bea
&&
-{1\over 2} L+ 2e^{\la} - \ap \nu' \phi' e^{-\la} +
{\ap \over 4} r^2\phi^{\pr 4}e^{-\la}=0, 
\label{4}\\
&&
\left[ \phi^{\se}+ \left({\nu'\over 2}-{\la'\over 2}-\phi'\right)
\phi' \right] \left[ -r^2 +\ap \left( e^{-\la}-1\right) \right]+
\nonumber \\
&&
+2r\left({\nu'\over 2}-{\la'\over 2}-\phi'\right)- \phi'
\left(2r+ \ap \la'  e^{-\la} \right) +2 -{L\over 2}=0, \\
&&
2r\nu^{\se}+2\ap e^{-\la}\left( \nu^{\se}\phi'+
\nu' \phi^{\se}- \nu'\phi'\la'\right)
-2r\left(\phi^{\pr 2}-\nu'\phi'\right)+\nonumber\\
&&
\left({\nu'\over 2}-{\la'\over 2}-\phi'\right)\left(-4r\phi' +4+
2r\nu'+2\ap\nu'\phi'e^{-\la} \right)-4r\phi^{\se}
 +{\ap\over 2} r \phi^{\pr
4} e^{-\la} =0,\\
&&
4r\phi'+2r^2\phi^{\se}-4-\nu'\left(2r+\ap\la' e^{-\la}\right)+
\nu^{\se}\left[-r^2+\ap
\left(e^{-\la}-1\right)\right]\nonumber\\
&&
-\ap e^{-\la}\left(-\la' r^2\phi^{\pr 3} +2r \phi^{\pr 3} +3 r^2
\phi^{\pr 2} \phi^{\se}\right) +L +\nonumber\\
&&
\left({\nu'\over 2}-{\la'\over
2}-\phi'\right)\left[r^2\left(2\phi'-\nu'\right)-4r+\ap\nu'
\left(e^{-\la}-1\right)-\ap r^2\phi^{\pr 3}e^{-\la} \right]=0,
\label{7} 
\eea
where
\beq
L(r)=r^2\phi^{\pr 2}-\phi'\left (r^2\nu'+4r\right)+2+
2r\nu'+2e^{\la} +\ap\nu'\phi' \left(e^{-\la}-1\right)-
{\ap\over 4}r^2\phi^{\pr 4} e^{-\la}.
\label {8}
\eeq
Of these four equations, only three are independent. The first
one, following from the variation of $\la$, does not contain
second derivatives and can be used as a constraint on the
initial conditions. 

For $\phi=$ const and $\ap=0$ the only non-trivial solution
of the above system is the well known singular Schwarzschild
metric, $e^\nu= e^{-\la} =1-2m/r$. For $\phi' \not=0$ and $\ap
\not=0$, on the contrary, there are non-trivial solutions with
no curvature singularities at the origin. We can easily check
that the above equations are indeed satisfied, at $r=0$, by 
\beq
\la(0)=0 , \,\,\,\, \,\la'(0)=0 , \,\,\,\,\, \nu'(0)=0 , \,\,\,\, \,
\nu^{\se}(0)=2/\ap , \,\,\,\,\, \phi'(0)={\rm const}\not=0 , 
\label{9}
\eeq
corresponding to a non-zero but finite value of the curvature
invariants at the origin. By using the Taylor expansion, and imposing the
initial conditions (\ref{9}), we then find a class of solutions
that around $r=0$ are approximated by
\bea
&&
\phi=\phi(0)+r\phi'(0)+{r^2\over 2} \phi^{\se}(0)+{\cal O}(r^3) ,
\nonumber\\
&&
\nu= \nu(0)+ {r^2\over \ap} +{\cal O}(r^3) ,
\,\,\,\,\,\,\,\,\,\,\,\,\,\,\,\,\,\,\,\, \la= -{r^2\over \ap} +
{\cal O}(r^3) 
\label{10}
\eea
where  $\phi(0)$ and $\nu(0)$ are left arbitrary, while $\phi'$
and $\phi^{\se}$ are determined by the field equations as 
(in units $\ap=1$): 
\beq
\phi'(0)= -1.414... ,\,\,\,\,\,\,\,\,\,\,\,\,\,\,\,\,\,
\phi^{\se}(0)= -0.4166...
\label{11}
\eeq
(there is also a non-trivial solution with $\phi^{\pr}(0)>0$, but
in that case the singularity appears at a finite distance from
the origin). 

It is interesting to note that, for small enough $r$, the metric
part of the solution (\ref{10}) approximates an anti-de Sitter 
background with 
\beq
e^{-\la}= 1+\La r^2 , \,\,\,\,\,\,\,\,\,\,\,\,\,\,\,\,\,\,
e^\nu= e^{\nu_0} \left(1+\La r^2\right), 
\label{12}
\eeq
and cosmological parameter $\La= 1/\ap$ (the constant
$\nu_0=\nu(0)$ can be absorbed by rescaling the time
coordinate). Such a metric parametrizes a maximally
symmetric manifold, whose constant curvature invariants are
determined by $\La$ as 
\beq
R_{\mu\nu\a\b}^2= 24 \La^2 , \,\,\,\,\,\,\,\,\,\,\,\,\,\,\,
R_{\mu\nu}^2= 36 \La^2 , \,\,\,\,\,\,\,\,\,\,\,\,\,\,\,
R^2= 144 \La^2 .
\label{13}
\eeq
The computation of the Ricci tensor around the origin, for the
approximated solution (\ref{10}), gives in fact $R_\mu^\nu(0)=
(3/\ap)\da_\mu^\nu$, in agreement with eq. (\ref{13}) for
$\La= 1/\ap$.

In the opposite limit $r\ra \infty$, the equations of motion 
(\ref{4})--(\ref{7}) are again asymptotically satisfied by a
maximally symmetric anti-de Sitter manifold, but with a
different value of the effective cosmological constant. By
setting
\beq
\phi=\phi_\infty -\ga \log r , \,\,\,\,\,\,\,\,\,\,\,\,\,
\nu=\nu_\infty +2 \log r , \,\,\,\,\,\,\,\,\,\,\,\,\,
\la=-\log \La -2 \log r  \,\,\,\,\,\,\,\,\,\,\,\,\,
\label{14}
\eeq
(where $\nu$ and $\la$ approximate the metric (\ref{12}) for
$\La r \gg1$), the field equations (\ref{4})--(\ref{7}) are in fact 
reduced, for $r\ra \infty$, to a system of four algebraic
equations:
\bea
-3-3\ga-{1\over 2} \ga^2+ 3\ap\La \ga +{3\over
8}\ap\La\ga^4 &=& 0 , 
\label{15}\\
3+2\ga+{1\over 2} \ga^2-2\ap\La \ga -\ap\La\ga^2
+{1\over 8}\ap\La\ga^4 &=& 0 ,  
\label{16}\\ 
12+8\ga +2 \ga^2-8\ap\La \ga -4\ap\La\ga^2
+{1\over 2}\ap\La\ga^4 &=& 0 ,  
\label{17}\\ 
-12 +6\ap\La -6\ga -\ga^2+3\ap\La \ga^3+{3\over 4}\ap\La
\ga^4 &=& 0 ,
\label{18}
\eea
which are not, of course, all independent (the left-hand side of
eqs. (\ref{16}) and (\ref{17}) are indeed proportional), and
which provide non-trivial solutions for the two unknown $\La$
and $\ga$.  Discarding  negative values of $\ga$, since we are
looking for configurations with decreasing dilaton, we are then left 
with two possible pairs of real solutions:
\bea
&&
\ap \La = 1.820... , \,\,\,\,\,\,\,\,\,\,\,\,\,\,\,\,\,\,\,\,\,
\ga=1.079... , \nonumber \\
&&
\ap \La = 2.194... , \,\,\,\,\,\,\,\,\,\,\,\,\,\,\,\,\,\,\,\,\,
\ga=0.8258... \,.
\label{19}
\eea

In the limit $r \ra \infty$ the solution (\ref{14}) satisfies the
condition of maximally symmetric manifold (eq. (\ref{13})),
with the asympotic value of the curvature fixed by the
numerical values (\ref{19}). The two constant parameters of the
solution, $\phi_\infty$ and $\nu_\infty$, can be chosen so as
to continuously match this anti-de Sitter configuration to the
other one approaching the origin. A possible example of a 
solution that is everywhere regular (with no event horizons) 
and smoothly interpolates between the two
asymptotic states of constant curvature, has beeen obtained
by integrating numerically eqs. (\ref{4})--(\ref{7}), and is
illustrated by the following four figures (plotted for $\ap=1$). 

In Fig. 1 we show the logarithmic derivative of the 
time and radial components of the
metric tensor, which display the typical anti-de Sitter
behaviour of eq. (\ref{12}), 
$\nu'=-\la'= 2\La r/(1+\La r^2)$, 
with the only difference that the
effective cosmological parameter $\La$ is slightly different in
the two regimes $r\ll \sqrt{\ap}$ and $r\gg \sqrt{\ap}$. 

In Fig. 2 we show that the string coupling $g_s=e^{\phi/2}$ is
decreasing for $r\ra \infty$, and that its first and second
derivatives are everywhere bounded. The normalization of
$\phi$ at $r=0$  is fixed by an arbitrary integration constant; 
it can always be chosen in such a way that $g_s<1$ 
everywhere. The figure corresponds to the particular case
$\phi(0)=-1$. 
\begin{figure}
\centerline{\epsfig{file=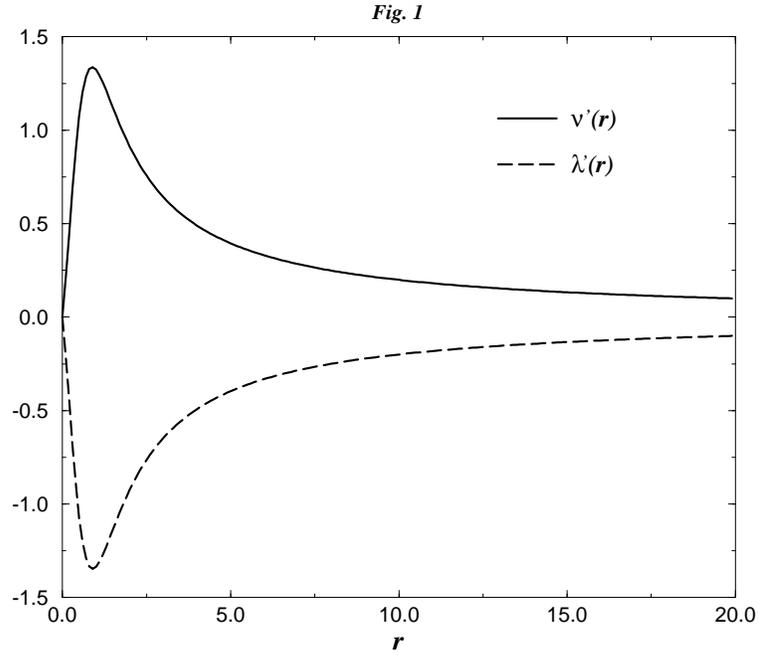,width=0.6\textwidth,angle=-90}}
\caption{\sl Radial behaviour of the logarithmic derivatives of the 
metric tensor.}
\end{figure}
\begin{figure}
\centerline{\epsfig{file=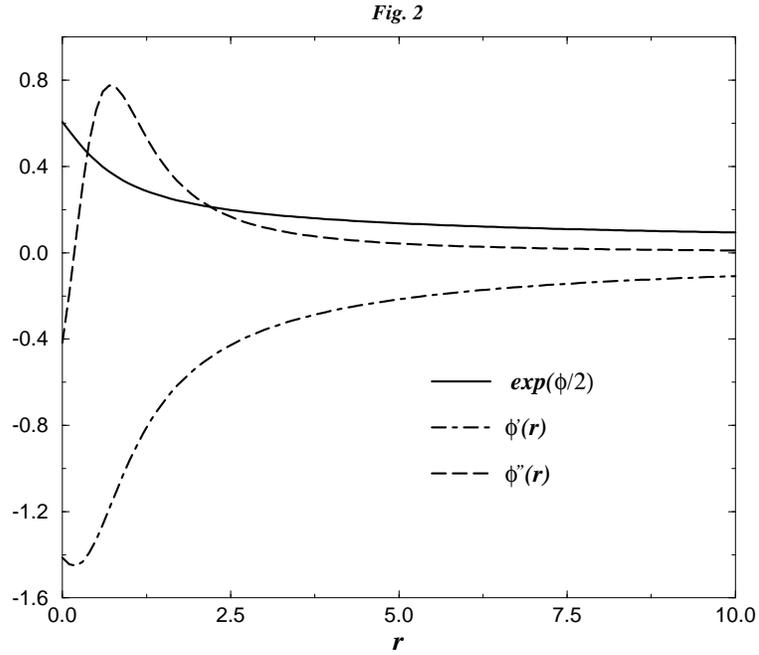,width=0.6\textwidth,angle=-90}}
\caption{\sl Radial behaviour of the string coupling $g_s=e^{\phi/2}$, 
and of the derivatives of the dilaton field.}
\end{figure}
\begin{figure}
\centerline{\epsfig{file=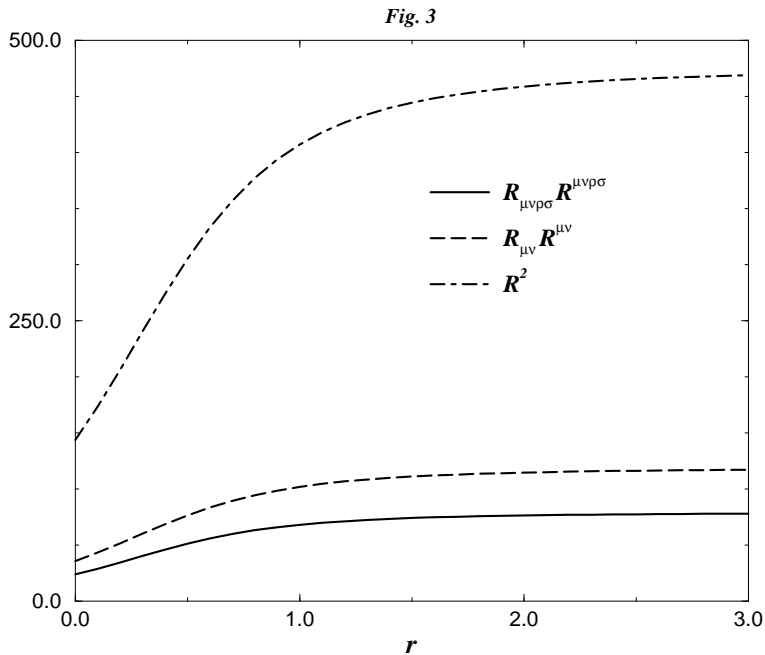,width=0.6\textwidth,angle=-90}}
\caption{\sl Radial behaviour of the curvature invariants.}
\end{figure}
\newline
In Fig. 3 we plot the curvature invariants
$R_{\mu\nu\a\b}^2$, $R_{\mu\nu}^2$ and $R^2$. Their radial
dependence interpolates between the constant values
determined, according to eq. (\ref{13}), by an effective
cosmological term $\La=1/\ap$ at $r \ra 0$ and $\La=1.82/\ap$
at $r \ra \infty$ (the latter corresponding to the first solution of eq.
(\ref{19})). 

In Fig. 4 we show that the ratios $R_{\mu\nu\a\b}^2
/R^2$ and $R_{\mu\nu}^2/R^2$ 
approach, asymptotically, the constant values $1/6=0.166$ and 
$1/4=0.25$ respectively, which are 
typical of maximally symmetric manifolds according to eq.
(\ref{13}). 
\begin{figure}
\centerline{\epsfig{file=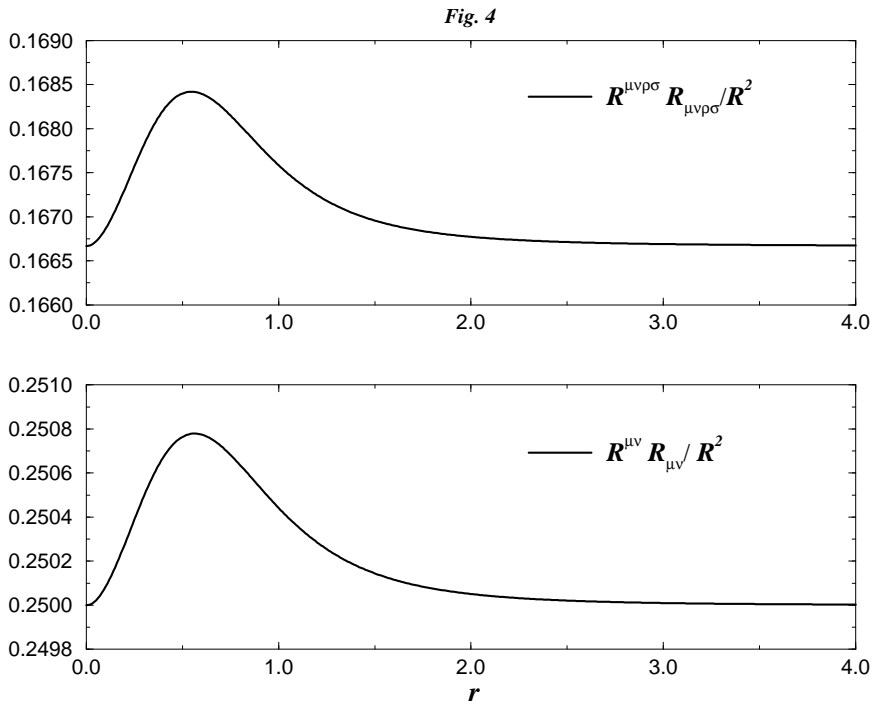,width=0.6\textwidth,angle=-90}}
\caption{\sl Radial interpolation between the two asymptotic anti-de 
Sitter configurations.}
\end{figure}

In conclusion, a few comments are in order. 
This class of backgrounds is certainly interesting as a class of
regular solutions of a higher-derivative model of gravity. For
what concerns string theory, however, it is presently unclear
whether such solutions can be extended to all orders in $\ap$,
to represent a non-trivial zero of the corresponding
sigma-model $\b$-functions. The truncation of the action  is
indeed unmotivated in a string theory context, when the
background reaches curvature scales of order one in string
units, like in the example studied in this paper. Also, as
discussed in the cosmological case \cite{6}, the existence of
solutions that are everywhere regular may be a property 
that depends on the particular field redefinition 
adopted, until the $\ap$ expansion of the string effective
action is truncated at a given finite order. This ambiguity can 
only be resolved by a solution corresponding to a true
conformal field theory, exact to all orders in $\ap$. The
investigation of this aspect of the problem is postponed to future works. 

Nevertheless, the existence of regular solutions in the
perturbative regime  of the quantum-loop
expansion ($g_s\ll1$), seems to support the
expectation that space-time singularities may be
regularized, already at a ``classical" level, by the finite-size 
``stringy" $\ap$ corrections.

It is important to stress, finally, that constant curvature
configurations may generally appear, asymptotically, when
higher-derivative terms are added to a lowest-order action.
Such configurations, however, are in general disconnected
from the origin by one (or more) curvature singularities, 
appearing at finite values of $r$. 
By contrast, for the action (\ref{1}) in which the higher-derivative
corrections appear precisely in the form dictated
by string theory, the regular asymptotic
regimes can be smoothly connected, and there are solutions in
which the  curvature is everywhere bounded. In this sense
the action (\ref{1}) automatically implements the
``limiting-curvature" hypothesis \cite{7}, often invoked to
regularize space-time singularities.

\vskip 2 cm
\noi
{\bf Acknowledgements}

\noi
We are grateful to Gabriele Veneziano for many useful conversations. A. 
B. and C. U. are partially supported by University of Pisa.

\end{document}